\documentclass[aps,pr,twocolumn,superscriptaddress]{revtex4-2}

\usepackage{mathtools}
\usepackage{graphicx}
\usepackage{mathrsfs}
\usepackage{amssymb}
\usepackage{xcolor}
\usepackage{amsmath}
\usepackage{bm}
\usepackage{amsmath}
\usepackage{physics}
\makeatletter\AtBeginDocument{\let\@elt\relax}\makeatother
\usepackage{amsmath}
\usepackage[normalem]{ulem}
\usepackage[unicode=true,colorlinks=true,citecolor=blue,urlcolor=blue]{hyperref}
\usepackage{tabularx}
\usepackage{ctable}

\def\be {\begin{equation}}
\def\ee {\end{equation}}
\def\bea {\begin{align}}
\def\eea {\end{align}}

\def\bee{\begin{eqnarray}}
\def\eee{\end{eqnarray}}
\def\BC {\begin{cases}}
\def\EC {\end{cases}}

\makeatother

\begin{document}
\title{Bleaching of the Terahertz Magneto-Photogalvanic Effect in CdHgTe Crystals with Kane Fermions }

\author{M. D. Moldavskaya}
\author{L. E. Golub}
\affiliation{Physics Department, University of Regensburg, 93040 Regensburg, Germany}

\author{V. V. Bel'kov}
\affiliation{Ioffe Institute, 194021 St. Petersburg, Russia}

\author{N.~N.~Mikhailov}
\author{S. A. Dvoretsky}
\affiliation{Rzhanov Institute of Semiconductor Physics, 630090 Novosibirsk, Russia}

\author{D.~Weiss}
\author{S.~D.~Ganichev}
\affiliation{Physics Department, University of Regensburg, 93040 Regensburg, Germany}

\begin{abstract}	
We report the observation and comprehensive study of the complex nonlinear intensity dependence of the magneto-photogalvanic effect (MPGE) current induced by terahertz (THz) radiation in Cd$_{x}$Hg$_{1-x}$Te films with inverted ($x = 0.15$) and  noninverted ($x = 0.22$) band structures.
The nonlinearities are studied for the resonant MPGE caused by cyclotron resonance, interband transitions between Landau levels, and ionization impurities, as well as for nonresonant indirect Drude-like optical transitions. We show that all these processes lead to the saturation of the photocurrent caused by absorption bleaching. We develop a theoretical framework for each mechanism,
which describes measured nonlinearities over an intensity range from 10$^{-3}$ to $4\times10^{4}$~W/cm$^2$. 
Furthermore, we demonstrate that the saturation processes for these absorption channels differ significantly, allowing us to analyze them independently by considering different intensity regimes. 
The observed nonlinearities enable us to determine the energy relaxation times of Kane fermions excited by CR and interband optical transitions.
\end{abstract}

\maketitle

\section{Introduction}
\label{intro}

In the past decade, studies of bulk CdHgTe crystals have attracted growing attention, because they host Kane fermions, which represent single valley relativistic states~\cite{Orlita2014}. 
The rest mass of Kane fermions changes sign at a critical temperature, while their velocity remains constant. Similar to the pseudospin Dirac-Weyl system, their energy dispersion relation features cones intersected at the vertex by an additional flat band. The energy dispersion asymptotically approaches a linear Weyl behavior, similar to relativistic electrons at high energies. Moreover, the narrow band gap allows for the study of both massive and massless fermions by appropriately selecting of experimental conditions~\cite{Orlita2014,Malcolm2015,Teppe2016,Rauch2017,Marcinkiewicz2017,Vasileva2020,Otteneder2020,Savchenko2020,Hubmann2020,Kazakov2021,Gadge2022,Krishtopenko2022,Moldavskaya2024}.
The  band structure of CdHgTe can be continuously tailored by adjusting the cadmium content or temperature. At a critical Cd concentration or temperature, the bandgap collapses as the system undergoes a semimetal-to-semiconductor topological phase transition between inverted and normal alignments. 
CdHgTe crystals are particularly  interesting because, 
when the band alignment is inverted -- realized at low Cd concentrations or low temperatures -- topologically protected surface states emerge.
In the presence of a magnetic field, a distinctive fan of Landau levels forms, giving rise to novel effects in magneto-optics. The advanced level of technological control and the unique properties of these materials,  which serve as condensed matter analogs of high-energy relativistic particles,  pave the way for the development of innovative electronic  devices. 

\begin{figure}[t]
	\centering \includegraphics[width=\linewidth]{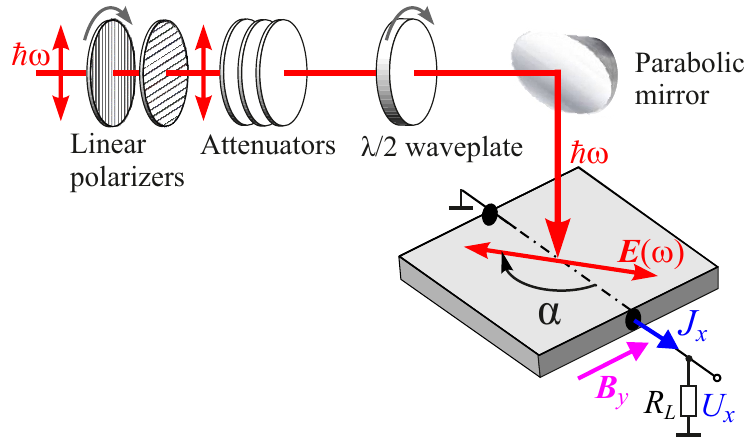}
	\caption{Schematic illustration of the experimental setup. A linearly polarized laser beam passes through two grating polarizers. Rotating the first polarizer while keeping the second polarizer fixed allows for a controlled variation of the radiation intensity at a defined polarization. Rotation of the lambda-half plate allows for variation of the angle $\alpha$, which defines the relative orientation of the radiation electric field vector with respect to the magnetic field direction.  The generated electric current $J_x$ is measured as the voltage drop $U_x$ across the load resistance $R_L=50$~Ohm.}
	\label{Fig_setup}
\end{figure}

In our recent work~\cite{Moldavskaya2024}, we reported on the observation and comprehensive study of the terahertz radiation induced multiple resonant and nonresonant magneto-photogalvanic effect (MPGE) in bulk Cd$_{x}$Hg$_{1-x}$Te films with Cd contents $x = 0.15$ and 0.22, subjected to an in-plane magnetic field. It has been shown that the nonresonant MPGE is caused by the Drude absorption. The observed multiple photocurrent resonances excited at different magnetic fields are caused by the CR, interband optical transitions involving Landau levels in the flat valence band, and impurity ionization.
All of the above results have been obtained at low irradiation intensities $I$, where the MPGE depends linearly on $I$.

Here we report the observation and detailed study of a complex intensity dependence of both resonant and non-resonant MPGE excited by intense linearly polarized terahertz radiation in Cd$_{x}$Hg$_{1-x}$Te films with Cd contents $x = 0.15$ and 0.22. 
The nonlinearity is studied in a wide range of intensities from mW/cm$^2$ to 45~kW/cm$^2$.
A developed microscopic theory of the nonlinear MPGE accurately fits all experimental data. We demonstrate that, over almost the entire intensity range, the MPGE nonlinearity is caused by the bleaching of the radiation absorption, characterized by different 
saturation intensities $I_s$ for different resonant and non-resonant absorption mechanisms. We show that the saturation of the nonresonant MPGE photocurrent is caused by the absorption saturation 
due to electron-gas heating. In contrast, the saturation of the resonant MPGE results from the slow relaxation of the photoexcited carriers back to the initial state of the direct optical transition, leading to CR, interband transitions between Landau levels, or impurity ionization. A drastic difference in the saturation intensities and low intensity magnitudes of the MPGE, caused by different absorption mechanisms, allowed us to fit the data across different irradiance ranges using only one fitting parameter while keeping the others fixed. From the fitting parameters and calculated absorption coefficients, we determined the energy relaxation times for the CR and interband absorption mechanisms.

\begin{figure}[t]
	\centering \includegraphics[width=0.8\linewidth]{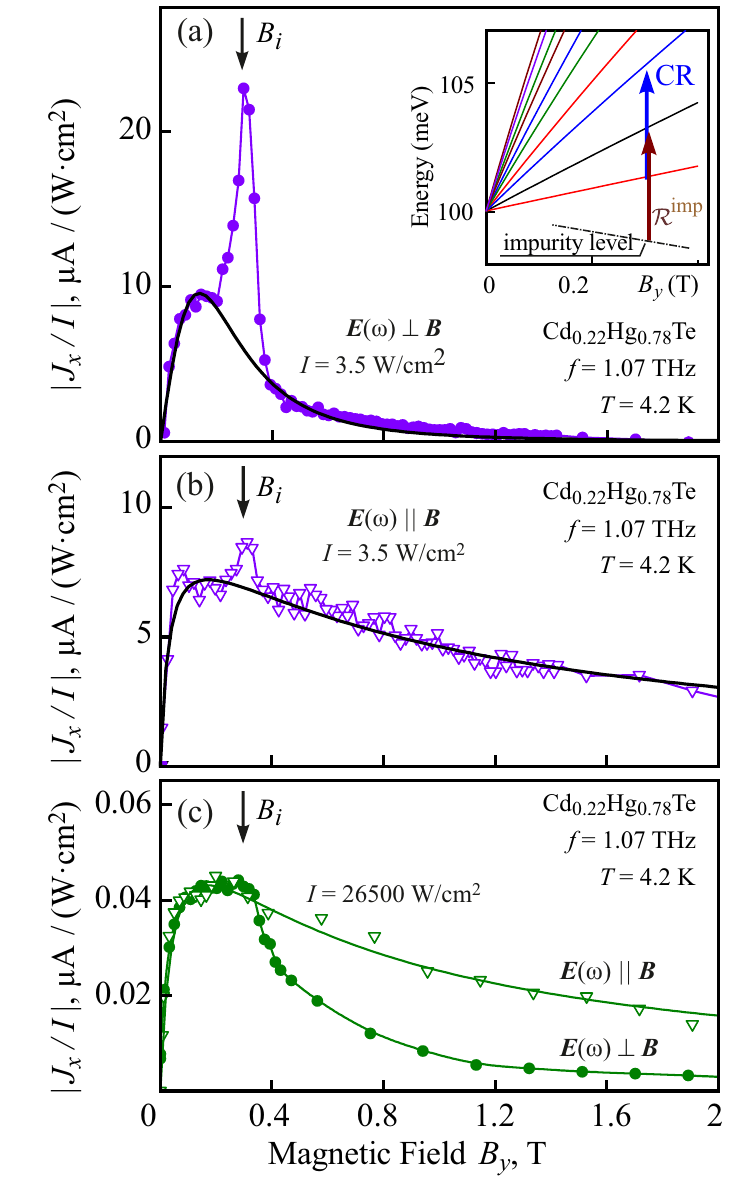}
	\caption{Magnetic field dependence of the photocurrent $J_x $($B_y$) normalized to the intensity radiation $I$ measured in sample B ($x$ = 0.22) at $T$ = 4.2 K.
		Data are shown for the radiation electric field both perpendicular and parallel to the magnetic field $\bm B$, and for two radiation intensities 3.5 W/cm$^2$ [panels (a) and (b)] and 26500 W/cm$^2$ [panel (c)].  Vertical down arrows indicate the magnetic field $B_i = 0.29$~T at which the dependence of the photocurrent on the radiation intensity is measured, see Fig.~\ref{Fig2_Int_022}. 
		The black solid curves in panels~(a) and~(b), and green solid curves in panel~(c), show the dependence of the Drude current according to the theory developed in Ref.~\cite{Moldavskaya2024}.  The inset in panel (a) shows the low magnetic field part of the energy spectrum of Cd$_{0.22}$Hg$_{0.78}$Te at $T=4.2$~K, calculated
		in Ref.~\cite{Moldavskaya2024} using the Kane model.
		Vertical arrows, whose lengths correspond to the photon energies used in our experiments $\hbar\omega =4.4$~meV ($f = 1.07$~THz), indicate resonant optical transitions at $B=B_i$. The arrows are labeled CR and ${\cal R}^{\rm imp}$, representing the polarization sensitive and polarization independent resonances, respectively.}
	\label{Fig2_B_022}
\end{figure}

\begin{figure}
\centering \includegraphics[width=\linewidth]{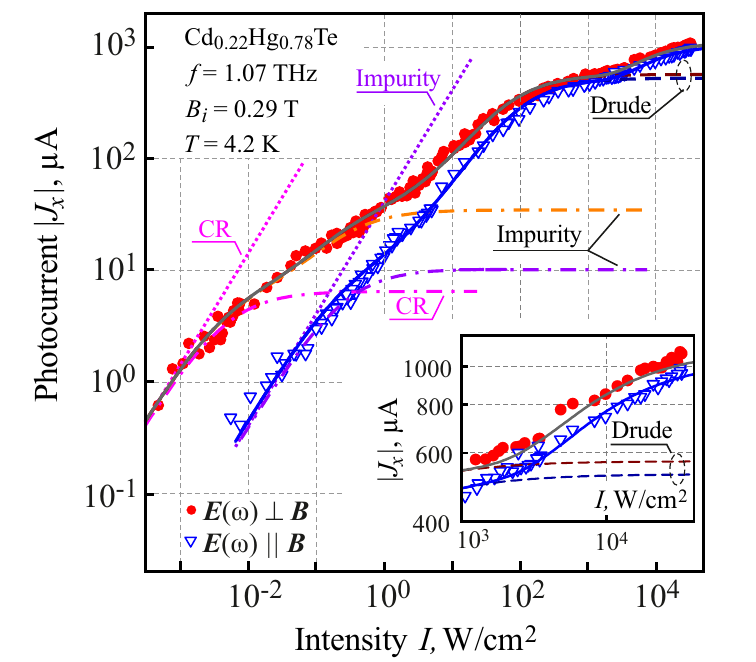}
	\caption{
	Intensity dependence of the photocurrent in sample B ($x=0.22$) obtained at $T$ = 4.2 K 
	and magnetic field $B_i = 0.29$~T.
	Data are shown for the radiation electric field perpendicular (solid dots) and parallel (open downward triangles) to the magnetic field $\bm B$. The inset 
	provides a zoomed-in view of the data at high intensities.
The fitting procedure is presented in Sec.~\ref{discussion} and the fitting parameters are given in Table~\ref{Tab2}. 
The dotted straight lines show the $J_x \propto I$ dependencies describing the data at the lowest intensity range.
Magenta dash-dotted, orange dash-dotted and brown dashed  curves are the 1st, 2nd and 3rd terms in Eq.~\eqref{CRA015}, respectively.  
The violet dash-dotted and navy dashed curves  are the 1st and 2nd terms in Eq.~\eqref{CRP015}.
Solid lines show fits of the data according to Eqs.~\eqref{impact},~\eqref{CRA015} and~\eqref{CRP015} obtained for the entire intensity range. 		 }
	\label{Fig2_Int_022}
\end{figure}

\begin{figure}[t]
\centering \includegraphics[width=\linewidth]{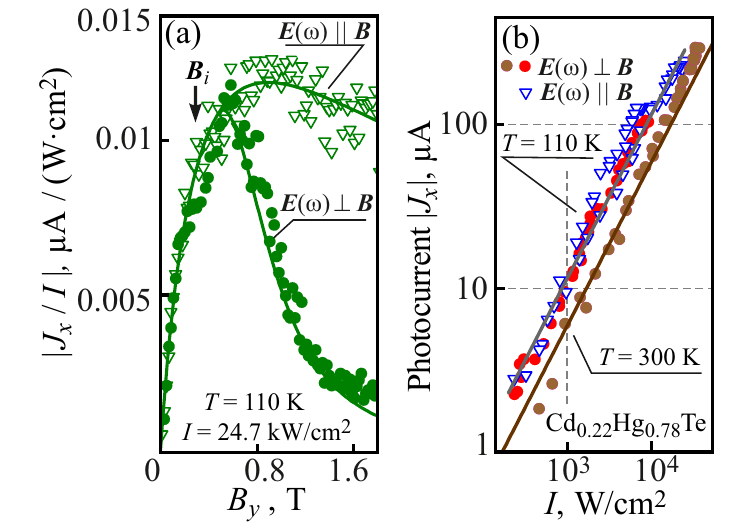}
	\caption{
	Panel (a): Magnetic field dependence of the  photocurrent $J_x(B_y)$, normalized to the intensity radiation $I$, measured in the sample B ($x$ = 0.22) at $T$ = 110 K.
		 Data are shown for both the radiation electric field perpendicular and parallel to the magnetic field $\bm B$ and radiation intensity $I$=24700 W/cm$^2$.  The arrow indicates the magnetic field position ($B_i=0.29$~T) at which the dependence of the photocurrent on the radiation intensity is measured.	
	The solid green curves show the dependence of the Drude current according to the theory developed in Ref.~\cite{Moldavskaya2024}.  Panel~(b): Intensity dependencies of the photocurrent  in sample B ($x$ = 0.22) obtained at $T$ = 110~K and 300~K for the
	magnetic field $B_i$.  Data are shown for the radiation electric field perpendicular and parallel to the magnetic field $\bm B$. Solid curves show the linear fit of the data. }
	\label{Fig_109K_022}
\end{figure}

\section{Samples and methods}

Experiments were carried out on two Cd$_{x}$Hg$_{1-x}$Te films with $x=0.15$ and $x=0.22$, which are representative for structures with inverted and normal band order. 
The samples were grown by standard molecular-beam epitaxy on (013)-oriented GaAs substrates, followed by a 30~nm ZnTe buffer layer and a 6~$\mu$m layer of CdTe. The film thickness with a constant Cd content $x$ was 4~$\mu$m for sample A ($x = 0.15$) and 7~$\mu$m for sample B ($x = 0.22$). 
The samples were square shaped with $5$~mm sides oriented along [110] and [1$\bar{1}$0]. Each sample was supplied with four indium ohmic contacts soldered to the edges centers, see Fig.~\ref{Fig_setup}. 
Details on the sample's composition, parameters as well as  transport and magnetotransport characteristics can be found in Refs.~\cite{Otteneder2020a,Moldavskaya2024}, where the same samples were studied. Most of the measurements presented here were carried out at liquid helium temperature. The magnetotransport and MPGE studies at low intensities have been performed in Ref.~\cite{Moldavskaya2024}. It was demonstrated that the MPGE is generated by the excitation of electrons with density and mobility $1.6\times10^{11}$\,cm$^{-2}$ and $2\times10^6$\,cm$^2$/V$\cdot$s for sample A and   $1.6\times10^{11}$\,cm$^{-2}$ and $10^5$\,cm$^2$/V$\cdot$s  for sample B, respectively.

\begin{figure}[t]
	\centering \includegraphics[width=\linewidth]{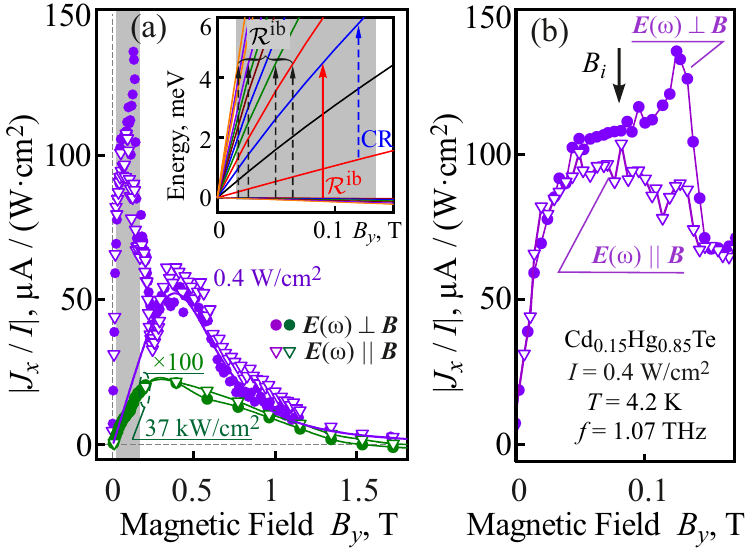}	
	\caption{
	Panel (a): Magnetic field dependence of the photocurrent $J_x $($B_y$) normalized to the intensity radiation $I$ measured in sample A ($x$ = 0.15) at $T$ = 4.2 K.
		Data are shown for the radiation electric field perpendicular (solid dots) and parallel (open downward triangles) to the magnetic field $\bm B$ and for two radiation intensities  
	0.4~W/cm$^2$ (violet symbols) and 37000~W/cm$^2$ (green symbols). 		
	The solid curves show the dependence of the Drude current according to the theory developed in Ref.~\cite{Moldavskaya2024}.
	Panel (b) shows a low-field zoom of the magnetic field dependence obtained at low intensity ($I=0.4$~W/cm$^2$). 
	 The downward-pointing arrow indicates the magnetic field $B_i =0.08$~T, at which the photocurrent dependence on radiation intensity is measured, see Fig.~\ref{Fig_Int_015}. The inset in panel (a) shows the low magnetic field part of the energy spectrum of Cd$_{0.15}$Hg$_{0.85}$Te, calculated for $T=4.2$~K  in Ref.~\cite{Moldavskaya2024} using the Kane model.
	 Vertical arrows, whose length corresponds to the photon energy used in our experiments $\hbar\omega =4.4$~meV ($f = 1.07$~THz), indicate resonant optical transitions at $B=B_i$ (solid arrow) and at other fields (dashed arrows). The arrows are labeled CR and ${\cal R}^{\rm ib}$, representing the polarization sensitive and polarization independent resonances, respectively.
	 {The inset reveals that a broad peak at fields $B\lesssim 0.14$~T is caused by multiple interband transitions. The  corresponding area is highlighted in gray in panel~(a) and the inset.}
	 }
	\label{Fig_B_015}
\end{figure}

\begin{figure}[t]
	\centering \includegraphics[width=0.9\linewidth]{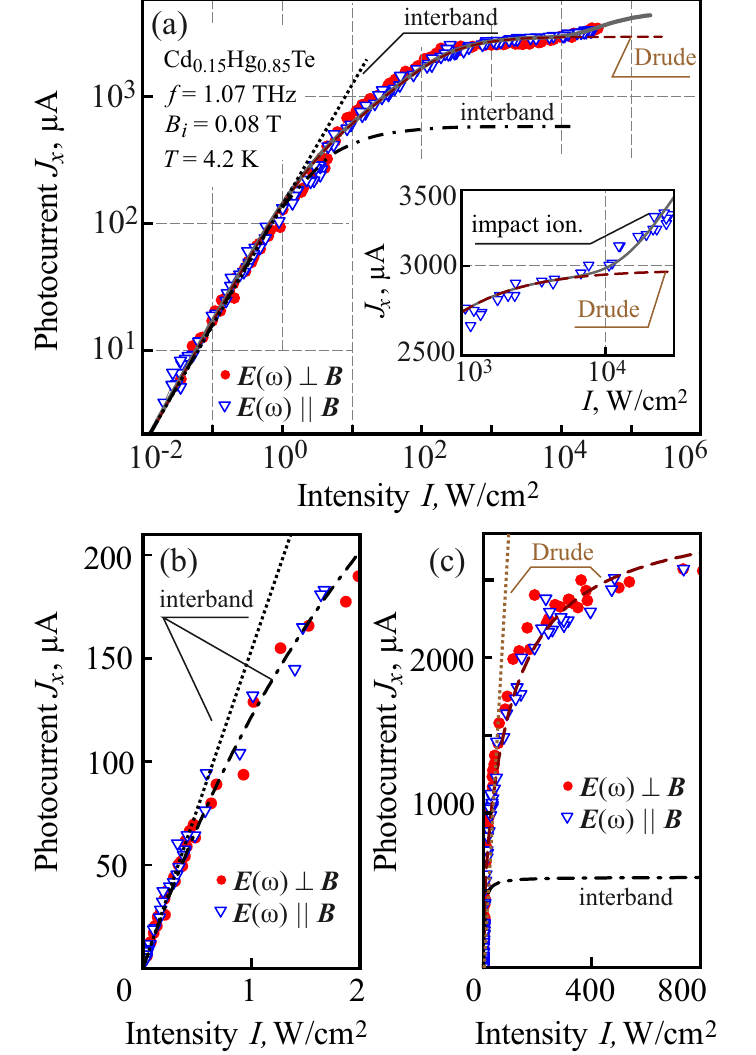}
	\caption{	Panel (a) shows the intensity dependence of the photocurrent in sample A ($x=0.15$), obtained at $T$ = 4.2 K for 
	magnetic field $B_i = 0.08$~T  over the entire  intensity range. Data are shown for the radiation electric field perpendicular (solid dots) and parallel (open downward triangles) to the magnetic field $\bm B$. Panels (b), (c) and the inset in panel (a) show the data for different intensity ranges.
	The fitting procedure is presented in Sec.~\ref{discussion}, and the fitting parameters are given in Table~\ref{Tab1}. 
The dotted straight lines show the $J_x \propto I$ dependence, describing the data in the lowest intensity range.
The black dash-dotted and brown dashed  curves correspond to the 1st and 2rd terms in Eq.~\eqref{015fit}, respectively.  
The solid line shows the fit  of the data according to Eqs.~\eqref{015fit} and~\eqref{impact} obtained for the entire intensity range. 
	}
	\label{Fig_Int_015}
\end{figure}

The samples were mounted in a temperature variable optical cryostat with $z$-cut crystal quartz windows and split coil superconducting magnet. To prevent the sample from being illuminated by residual infrared radiation ($\lambda < 10$~$\mu$m), we used TPX (4-methyl-penten) and Teflon plates, and all windows of the cryostat were always covered with a thick film of black polyethylene. The measurements were performed in the Voigt geometry using normally incident linearly polarized radiation and an in-plane magnetic field up to 2~T in $y$-direction, see Fig.~\ref{Fig_setup}.  For additional room temperature  measurement a magnetic field of up to 1~T was generated by an electromagnet. The samples were irradiated by a high-power pulsed molecular  laser optically pumped by a line-tunable, transversely excited atmospheric pressure (TEA) CO$_2$ laser~\cite{Ganichev1982,Ganichev2005}. The laser, with NH$_3$ gas as active medium, provided linearly polarized radiation at $f = 1.07$~THz ($\lambda = 280$~$\mu$m, $\hbar \omega =4.4$~meV). The molecular laser generated single pulses with a duration of about 100~ns and a repetition rate of 1~Hz. The beam position and profile were controlled using a pyroelectric camera. The almost Gaussian-shaped beam was focused onto the sample using an off-axis parabolic mirror, creating a spot with an area of 0.16~cm$^2$. The radiation pulses were monitored during the measurements using a terahertz photon-drag detector made of $n$-type Ge crystal~\cite{Ganichev2006,Ganichev1985}.  The maximum radiation power was approximately 7~kW, and the radiation intensity about 45~kW/cm$^2$.

The photocurrent $J_x$ was measured in unbiased structures from the voltage drop $U_x$ across a load resistor $R_L = 50$~Ohm in a closed circuit configuration. The voltage was recorded with a storage oscilloscope and the photocurrent was calculated according to $J_x = U_x/R_\parallel$, where $R_\parallel = R_sR_L/(R_s + R_L)$ and $R_s$ is the sample resistance. To vary the angle $\alpha$ between the electric field of the linearly polarized radiation  $\bm{E}$ and the magnetic field $\bm{B}$ we used a crystalline quartz $\lambda/2$ plate, see Fig.~\ref{Fig_setup}. 
The photocurrent was studied in two configurations, the cyclotron resonance active
(CRA, $\bm E \perp \bm B$, $\alpha = 0$) and the cyclotron resonance passive 
(CRP, $\bm E \parallel \bm B$, $\alpha = 90^\circ$). 

To vary the intensity of the laser radiation over more than seven orders of magnitude, from 1 mW/cm$^{2}$ to 45~kW/cm$^{2}$, we used both, a crossed polarizer setup consisting of two wire grid polarizers~\cite{Hubmann2019, Candussio2021a},  and calibrated Teflon attenuators, see Fig.~\ref{Fig_setup}. In the former case, the linearly polarized laser radiation first passed through the wire grating polarizer. The rotation of this polarizer resulted in a decrease in radiation intensity and a rotation of the polarization state. The radiation then passed through a second wire polarizer, which was  fixed in position. This further reduced the radiation intensity and ensured that it remained consistently polarized. The attenuators were applied to further reduce  the radiation intensity down to 1~mW/cm$^{2}$. Since the samples were extremely sensitive to THz radiation, special absorbing screens were used to prevent any photoresponse of the sample caused by THz radiation reflected or diffracted by any elements of the setup. The screens ensured that the signal was zero when a metal plate was placed in the beam path.

\section{Experimental results}

\subsection{Results for the structures with the noninverted band order ($x=0.22$)}
\label{Results022}

Figure~\ref{Fig2_B_022} shows the magnetic field dependencies obtained in sample B ($x=0.22$) at $T=4.2$~K for low ($I=3.5$~W/cm$^2$) and high ($I=26.5\times 10^3$~W/cm$^2$) radiation intensities. 
The photocurrent depends nonmonotonously on the magnetic field exhibiting a broad maximum at ${B \approx 0.2}$~T, see black solid lines in Fig.~\ref{Fig2_B_022}(a) and Fig.~\ref{Fig2_B_022}(b) as well as green solid lines in Fig.~\ref{Fig2_B_022}(c). Furthermore, at low intensities,  it is resonantly enhanced at $B=0.29$~T. While in the CRA configuration ($\alpha = 0$), see Fig.~\ref{Fig2_B_022}(a), the resonance is strong, in the CRP configuration ($\alpha = 90^\circ$), it becomes rather weak, but remains clearly detectable, see Fig.~\ref{Fig2_B_022}(b).
The origin of all these photocurrents is well understood and described in detail in our previous publication, see Ref.~\cite{Moldavskaya2024}. It has been demonstrated that the nonmonotonic nonresonant photocurrent arises from the MPGE caused by indirect Drude-like optical transitions. It exhibits qualitatively similar magnetic field dependencies.
While at low $B$ this signal is almost the same  for CRA and CRP, at $B \gtrsim 0.4$~T they have different magnitudes for $\alpha = 90^\circ$ and $\alpha = 0$, which is attributed to the characteristic polarization dependence of the MPGE~\cite{Foot_pol_dep}.
The analysis performed in Ref.~\cite{Moldavskaya2024} also demonstrates that the polarization independent resonant MPGE (${\cal R}^{\rm imp}$) is caused by the impurity ionization, while the resonant MPGE detected in CRA configuration is generated by the cyclotron resonance, see the inset in Fig.~\ref{Fig2_B_022}(a). 

An increase of the radiation intensity by five orders of magnitudes quantitatively alters the magnetic field dependence. Figure~\ref{Fig2_B_022}(c) shows that at high intensities, both  resonant photocurrents vanish, leaving only the nonmonotonic photocurrent. To understand this behavior, we studied the intensity dependence of the MPGE at fixed magnetic field $B_i=0.29$~T, which corresponds to the superpositions of resonant and nonresonant photocurrents, see inset to Fig.~\ref{Fig2_B_022}(a). Figure~\ref{Fig2_Int_022} shows the intensity  dependencies obtained for the CRP ($\alpha = 90^\circ$) and CRA ($\alpha = 0$) configurations. The plot demonstrates that the photocurrent exhibits a complex intensity dependence, showing sublinear behavior over almost the entire intensity range.
In Sec.~\ref{discussion} we analyze this dependence and show that  it is caused by the saturation of the radiation absorption with different photocurrent magnitudes and saturation parameters for different absorption mechanisms.

At high temperatures all resonances in the magnetic field disappear and the photocurrent exhibits a smooth nonmonotonic behavior for both configurations, see Fig.~\ref{Fig_109K_022}(a) for $T=110$~K. The figure reveals that at low magnetic field the nonresonant MPGE for $\alpha =0$ and $90^\circ$ has the same magnitudes, i.e., it is dominated by the polarization independent mechanism, which is addressed in Ref.~\cite{Foot_pol_dep}. Similar to low $T$ results, at $B \gtrsim 0.5$~T  
the polarization dependent and independent contributions become comparable, see Fig.~\ref{Fig_109K_022}(a).
Also intensity dependence changes at high temperature. Figure~\ref{Fig_109K_022}(b) demonstrates that at high $T$ the MPGE photocurrent  depends linearly on $I$ in the  whole range of  radiation intensities.

\subsection{Results for the structures with the inverted band order ($x=0.15$ and $T=4.2$~K)}

We now describe the results for the structures with inverted band order, which is realized in sample A ($x=0.15$) by cooling the sample down to liquid helium temperature. Figure~\ref{Fig_B_015} shows the magnetic field dependencies obtained for low ($I=0.4$~W/cm$^2$) and high ($I=37\times 10^3$~W/cm$^2$) radiation intensities. At low intensities, the photocurrent is described by a superposition of  the nonmonotonic component, exhibiting a maximum at $B\approx 0.4$~T, and two resonant contributions: a sharp one with a maximum at $B=0.11$~T and a broad one with a maximum at lower fields, see Fig.~\ref{Fig_B_015}(b). 
The analysis performed in Ref.~\cite{Moldavskaya2024} demonstrated that the nonresonant photocurrent (maximum at $B\approx 0.4$~T) is caused by the MPGE due to Drude-like absorption.
The sharp resonance (at $B=0.11$~T) detected only in the CRA configuration is caused by the CR-induced MPGE, see Fig.~\ref{Fig_B_015}(b). Additionally, the broad resonance at low magnetic field, detected in both CRA and CRP configurations are
caused by a superposition of several resonant optical transitions between LLs in the almost flat valence band and different LLs in the  conduction bands, denoted as ${\cal R}^{\rm ib}$ in the inset of Fig.~\ref{Fig_B_015}(a). Similar to the sample with the normal band ordering, at high radiation intensities the resonances are not detected and the photocurrent shows nonmonotonic $B$-field dependence with a maximum at $B\approx 0.3$~T, see green symbols in Fig.~\ref{Fig_B_015}(a).

Studying an intensity behavior of the photocurrent detected in sample A ($x=0.15$) at $T=4.2$~K in CRA and CRP configurations, we observed, similar to sample B ($x=0.22$), a complex behavior dominated by saturation with increasing radiation intensity.
Figure~\ref{Fig_Int_015} presents the intensity dependence obtained for a magnetic field $B_i=0.08$~T, at which the MPGE is caused by the interband transitions between Landau levels in the valence and conduction bands
superimposed by the MPGE due to Drude-like absorption, see the inset in Fig.~\ref{Fig_B_015}(a). The analysis of the observed dependencies is presented in the next section.

The intensity dependence of the photocurrent excited in the sample at higher temperature is beyond the scope of this study, as in this case the band order changes to the non-inverted one, which corresponds to the situation described in Sec.~\ref{Results022}.

\section{Theory}

 Now we analyze the observed highly nonlinear behavior of the MPGE current.  Below, in Sec.~\ref{discussion}, we show that it is caused by the saturation of radiation absorption due to different mechanisms, including cyclotron resonance, inter-band transitions between the Landau levels in the valence and conduction bands, photoionization of impurity levels and Drude-like intraband absorption. 
 Indeed, the MPGE current is given by
 \begin{equation}
 \label{J_K}
J =I \sum_{i}A^{i}  \mathcal K^{i}(I),
\end{equation}
where $i$ enumerates various absorption channels, $\mathcal K^{i}$ is the corresponding absorption coefficient, which depends on the radiation intensity, and the factors $A^{i}$ are determined by the dominant MPGE mechanisms~\cite{Moldavskaya2024}.

 We demonstrate in Sec.~\ref{discussion} that the entire complex behavior of the MPGE current, as the intensity varies over eight orders of magnitude, can be well described by the superposition of MPGE from different absorption channels, each of which saturates at substantially different intensities. To carry out  this analysis, we  derive the nonlinear absorption coefficient for each absorption channel addressed above.

\subsection{Bleaching of two-level systems}

First, we consider a model of absorption bleaching in two-level systems, which can be used to describe cyclotron resonance, inter-band transitions between the Landau levels in the valence and conduction bands, and the photoionization of impurity levels.

Let us analyze a two-level system absorbing radiation with photon energy $\hbar\omega=E_2-E_1$, where $E_{1,2}$ are the energies of the two levels. At low intensity $I$, let the optical transition rate $1 \to 2$ be $W \propto I$. Then, at arbitrary intensity, the nonequilibrium steady-state electron concentrations at the levels $n_{1,2}$ satisfy the following rate equations:
\begin{equation}
	W(n_1-n_2)= - {n_1 - n_1^{(0)} \over \tau_1}, \quad
	W(n_1-n_2)= {n_2 - n_2^{(0)} \over \tau_2}.
\end{equation}
Here $n_{1,2}^{(0)}$ are the equilibrium concentrations in the absence of radiation, and $\tau_{1,2}$ are the relaxation times, which describe the processes of equilibration of the level occupations. 

Solving this system, we obtain for the concentration difference
\begin{equation}
	n_1-n_2 = {n_1^{(0)}-n_2^{(0)} \over 1+ 2W\tau}, \quad \tau = {\tau_1 + \tau_2\over 2}.
\end{equation}

The absorbed power at arbitrary intensity is given by
\begin{equation}
	{\mathcal K I\over \hbar \omega} = W (n_1-n_2),
\end{equation}
where $\mathcal K$ is the absorption coefficient. Therefore we get for the intensity dependence of $\mathcal K$:
\begin{equation}
	\label{K}
	\mathcal K(I) = {\mathcal K_0 \over 1+ I/I_s},
\end{equation}
where $\mathcal K_0=W\qty(n_1^{(0)}-n_2^{(0)})\hbar\omega /I$ is the low-intensity value of $\mathcal K$, and the saturation intensity reads
\begin{equation}
	\label{Is}
 I_s = {I \over 2\tau W} = {\hbar\omega \qty(n_1^{(0)}-n_2^{(0)}) \over 2\tau\mathcal K_0 }.
\end{equation}

Below we obtain $\mathcal K_0$ and $I_s$ for different absorption mechanisms. To be specific, we label them as 'CR' for cyclotron resonance, 'ib' for inter-band transitions between the Landau levels in the valence and conduction bands, and 'imp' for the photoionization of impurity levels.

\subsubsection{Saturation at CR}

Under the conditions of CR transitions between two Landau levels, the low-intensity absorption coefficient, which appears in Eq.\eqref{K} is given by
\begin{equation}
	\mathcal K_0^{\rm CR} = {4Z_0 e^2 \tau_p \qty(n_1^{(0)}-n_2^{(0)}) \over m  n_\omega [1+(\omega \pm \omega_c)^2\tau_p^2] },
\end{equation}
where two signs correspond to the passive (P) and active (A) polarizations, $m$ and $\tau_p$ are the effective mass and the momentum relaxation time in the conduction band, and $n_\omega$ is the refractive index at the frequency $\omega$.
According to Eq.~\eqref{Is}, this yields for the saturation intensity at CR:
\begin{equation}
	I_s^{\rm CR} ={m n_\omega \hbar \omega [1+(\omega \pm \omega_c)^2\tau_p^2] \over 8Z_0 e^2 \tau^{\rm CR} \tau_p }.
\end{equation}
Here, $\tau^{\rm CR}$ is the time for the carrier to return to the initial state of the optical transition.
We see that, for the active polarization, it takes the form
\begin{equation}
\label{IsCR_A}
	I_s^{\rm CRA} ={m n_\omega  \hbar \omega \over 8Z_0 e^2 \tau^{\rm CR} \tau_p },
\end{equation}
and for the passive one
\begin{equation}
I_s^{\rm CRP} =(2\omega_c\tau_p)^2 I_s^{\rm CR,A} \gg I_s^{\rm CR,A} .
\end{equation}

\subsubsection{Saturation at interband transitions between the valence- and conduction-band Landau levels}

Now we consider direct optical transitions in quantizing magnetic fields. We assume that a motion along the magnetic field is unimportant (this is supported by an argument that the maximal 1D density of states is at the band edges).
Then we have resonant optical transitions between two levels and apply the above formalism.

The probability of the direct optical transition between the Landau levels is given by the Fermi's golden rule. Taking into account a width of the transition $\hbar/\tau^{\rm ib}$, we get 
\begin{equation}
	W = {2\tau^{\rm ib}\over \hbar^2}\abs{P_{\rm ib}E}^2,
\end{equation}
where $P_{\rm ib}$ is the dipole matrix element of the transition, and $E$ is the radiation electric field amplitude related to the intensity by 
$I=n_\omega\abs{E}^2/(2Z_0)$, where $Z_0$ is the vacuum impedance.
Therefore, we obtain from Eq.~\eqref{Is}
\begin{equation}
\label{Is_ib}
	I_s^{\rm ib} ={n_\omega \hbar^2 \over 8 Z_0  (\tau^{\rm ib}\abs{P_{\rm ib}})^2 }.
\end{equation}

\subsubsection{Saturation at impurity photoionization}

For the optical transitions from the impurity levels to the conduction band, we have the probability $W$ in a form similar to the interband transitions:
\begin{equation}
	W = {2\tau^{\rm imp}\over \hbar^2}\abs{P_{\rm imp}E}^2,
\end{equation}
where $P_{\rm imp}$ is the corresponding matrix element and the relaxation time $\tau^{\rm imp}$ is the capture time to the impurity states. Then  Eq.~\eqref{Is} takes the form
\begin{equation}
	I_s^{\rm imp} ={n_\omega \hbar^2 \over 8Z_0  (\tau^{\rm imp}\abs{P_{\rm imp}})^2 }.
\end{equation}

\subsection{Bleaching at intraband absorption}

Finally, we derive the saturation of the intraband (Drude-like) absorption. It occurs due to a dependence of the low-intensity absorption coefficient $\mathcal K_0$ on electron temperature -- the bolometric effect. This dependence naturally arises from the factor
\begin{equation}
	\mathcal K_0^{\rm Dr}(T) \propto \sum_{\bm k} \qty[f_0(\varepsilon_k)-f_0(\varepsilon_k+\hbar\omega)],
\end{equation}
where $\varepsilon_k$ is the energy dispersion in the band, and $f_0(\varepsilon)$ is the equilibrium distribution.
At $\omega\tau_p \gg 1$ and nondegenerate statistics this  yields 
\begin{equation}
	\label{K_0_T}
	\mathcal K_0^{\rm Dr, nondeg}(T) = {4 Z_0 e^2 N\over m  n_\omega \omega^2 \tau_p} \qty[1-{\rm e}^{-\hbar\omega/(k_{\rm B}T)}], 
\end{equation}
where $N$ is the electron concentration and $k_{\rm B}$ is the Boltzmann constant.

Taking into account this temperature dependence and assuming that the absorption results in the heating of the electron gas, leading to the establishment of an equilibrium distribution with a higher electron temperature $T_e=T+\Delta T$, we obtain for the intensity dependent absorption coefficient $\mathcal K^{\rm Dr}(I)$:
\begin{equation}
	\mathcal K^{\rm Dr} = \mathcal K_0^{\rm Dr}(T) + \pdv{\mathcal K_0^{\rm Dr}}{T}\Delta T, \quad N {k_{\rm B}\Delta T \over \tau_\varepsilon}=\mathcal K^{\rm Dr} I.
\end{equation}
Here $\tau_\varepsilon$ is the energy relaxation time in the electron gas.
This system of equations again yields
\begin{equation}
	\label{KDr}
	\mathcal K^{\rm Dr}(I) = {\mathcal K_0^{\rm Dr} \over 1+ I/I_s^{\rm Dr}}, 
\end{equation}
where
\begin{equation}
	\label{Is_Drude}
 I_s^{\rm Dr} = {N k_{\rm B} \over \tau_\varepsilon \abs{\partial \mathcal K_0^{\rm Dr}/\partial T}}.
\end{equation}
Here we took into account that $\partial \mathcal K_0^{\rm Dr}/\partial T<0$.
For the nondegenerate statistics, we have from  Eq.~\eqref{K_0_T}:
\begin{equation}
	I_s^{\rm Dr,nondeg} = {m  n_\omega   \omega \tau_p\over 4Z_0\hbar e^2 \tau_\varepsilon   } (k_{\rm B}T)^2 \qty[{\rm e}^{\hbar\omega/(k_{\rm B}T)} - 1].
\end{equation}

\section{Analysis of the MPGE nonlinearity}
\label{discussion}

\subsection{Structures with the inverted band order ($x=0.15$ and $T=4.2$~K)}

The intensity dependence of the MPGE current in the structures with inverted band order has been obtained in sample A ($x=0.15$) at $T=4.2$~K under a magnetic field {$B_i=0.08$~T}. 
The inset in Fig.~\ref{Fig_B_015} shows that 
the resonant MPGE current is caused by 
the interband transitions between Landau levels in the valence and conduction bands. This MPGE is superimposed by the intraband indirect (Drude-like) mechanism, which contributes across the entire magnetic field range.
Both types of MPGE can be excited in the  CRA  ($\alpha = 0$) and CRP ($\alpha = 90^\circ$) configurations. Figure~\ref{Fig_Int_015} shows that for  {$B_i=0.08$~T} the data for these configurations almost coincide in the whole intensity range.
As obtained in the previous section, at high intensities, both types of photocurrents $J \propto I\mathcal K(I)$, Eq.~\eqref{J_K}, saturate according to Eqs.~\eqref{K} and~\eqref{KDr}, with the saturation intensities given by Eqs.~\eqref{Is_ib} and~\eqref{Is_Drude}, respectively. Consequently, the total current is described by
\begin{equation}
	\label{015fit}
J_x^{\rm MPGE} = {A^{\rm ib}I\over 1+ I/I_{s}^{\rm ib}} + {A^{\rm Dr}I\over 1+ I/I_{s}^{\rm Dr}}.
\end{equation}
Here, $A^{\rm ib}$ and $A^{\rm Dr}$  are the slopes of the MPGE current intensity dependencies in the linear regime for the interband and Drude optical transitions, respectively, and $I_s^{\rm ib}$, $I_s^{\rm Dr}$ are the corresponding saturation intensities.
Figure~\ref{Fig_Int_015}(a) displays the data and the fit according to Eq.~\eqref{015fit} on a double logarithmic scale. This demonstrates that the above equations accurately describe all experimental data obtained within the intensity range from 0.01 to 10000~W/cm$^2$, covering six orders of magnitude. The corresponding fitting parameters are given in Table~\ref{Tab1}. The higher intensity range is discussed below. While Eq.~\eqref{015fit} has four fitting parameters, we leveraged the drastic difference between the amplitudes and saturation intensities of the interband and Drude processes. By analyzing different intensity ranges, we achieved  a fit by varying only one parameter while keeping the others constant. 

Figure~\ref{Fig_Int_015}(b) shows the data for the low intensity range from 0.01 to 2~W/cm$^2$, displayed on a double linear scale. At intensities up to about 0.3~W/cm$^2$, the MPGE current increases linearly with increasing $I$, as indicated by the black dotted line. From the slope we obtained $A^{\rm ib} = 154$~$\mu$A/(W/cm$^2$). At higher intensities, the current saturates according to  Eq.~\eqref{K} with a saturation intensity $I_{s}^{\rm ib} = 3.8$~W/cm$^2$, see dash-dotted black line in Fig.~\ref{Fig_Int_015}(b) and  Fig.~\ref{Fig_Int_015}(a)~\cite{FOOTPULSE}. 
The latter shows that at higher intensities ($I \gtrsim 2$~W/cm$^2$) the MPGE current continues to grow, whereas the calculated current [black dotted-dashed curves in Figs.~\ref{Fig_Int_015}(a) and (c)] becomes fully saturated. The further growth of the MPGE current is due to dominance of the Drude-like contribution over the fully saturated component originating from interband transitions.
Next, we fit the data in the intensity range from 2~W/cm$^2$ up to approximately $20$~W/cm$^2$ using the previously defined $A^{\rm ib}$ and $I_{s}^{\rm ib}$. In this range, the current exhibits an almost linear dependence shown by the brown dotted line in Fig.~\ref{Fig_Int_015}(c).
From this, we obtain the next fitting parameter, $A^{\rm Dr}$, which is approximately 7~times larger than the $A^{\rm ib}$, see Table~\ref{Tab1}. A further increase in intensity leads to the saturation of the MPGE due to the Drude absorption, as shown by the
brown dashed lines in Fig.~\ref{Fig_Int_015}.
Extending the analyzed intensity range to 10000~W/cm$^2$, we obtain $I_{s}^{\rm Dr} \gg I_{s}^{\rm ib}$, see 
Table~\ref{Tab1}.

\begin{table}[t]
	\centering
	\caption{Parameters used for fitting the data for sample A ($x=0.15$) in Fig.~\ref{Fig_Int_015} according to Eqs.~\eqref{015fit} and \eqref{impact}. 
		The fitting parameters are given for the interband and Drude mechanisms of the MPGE. 
		}
		\begin{tabularx}{0.5\textwidth}{XXXXXX}
			\toprule[0.05cm]\addlinespace[0.2cm]
			         & Transitions   & A$^i$ ($\mu$A) & $I_s^i$ (W/cm$^2$)   \\
			\midrule[0.025cm]\addlinespace[0.1cm]
			CRA, CRP & Interband     &   154          &    3.8               \\
			\addlinespace[0.1cm]\bottomrule[0.05cm]
		\end{tabularx}
		\label{Tab1}
	\end{table}

In summary, our analysis demonstrates that the bleaching of the MPGE current, observed over the intensity range from  0.01 to 10000~W/cm$^2$, can be explained by the saturation of two absorption channels: direct interband and indirect Drude-like optical transitions. The fit based on Eq.~\eqref{015fit}, which accounts for both processes , accurately describes all data for $I<$ 10000~W/cm$^2$, see the  dashed brown lines in Fig.~\ref{Fig_Int_015}(a). The amplitudes of the MPGE due to the interband transitions are approximately an order of magnitude larger than those caused by the Drude-related absorption, see Table~\ref{Tab1}. This is the expected result, taking into account that the absorption coefficient for the former mechanism is  is much higher than for the latter. Consequently, the MPGE current caused by interband transitions dominates the photocurrent at low intensities. The analysis also reveals that $I_{s}^{\rm ib} \ll I_{s}^{\rm Dr} $. Thus at higher intensities, the interband photocurrent totally saturates, and the Drude one becomes dominant, see brown dashed lines in Fig.~\ref{Fig_Int_015}. This difference in the saturation intensities also explains that at high intensities, no resonances are detected and the magnetic field dependence is described by a smooth, nonmonotonic behavior attributed to the Drude mechanism, see the green symbols in Fig.~\ref{Fig_B_015}(a). The difference of the saturation intensities $I_{s}^{\rm ib}$ and $I_{s}^{\rm Dr}$ is not surprising. Indeed, Eqs.~\eqref{Is},~\eqref{Is_ib}, and~\eqref{Is_Drude} demonstrate that the saturation intensities are proportional to the reciprocal absorption coefficients and reciprocal energy relaxation times, both of which are larger for the Drude process, which is  characterized by a smaller absorption coefficient and faster intraband relaxation. 

Our data allow us to determine  the interband relaxation time $\tau^{\rm ib}$ 
using Eq.~\eqref{Is_ib}. The dipole transition matrix element can be estimated as $P_{\rm ib} \approx ev/\omega$, where $v$ is the Kane fermion velocity. For $v=10^8$~cm/s~\cite{Teppe2016}, $\omega/(2\pi)=1$~THz
and $I_s^{\rm ib} =3.8$~W/cm$^2$, we get an estimate of $\tau^{\rm ib} \approx 1.3$~ps.
Notably, similar values have been detected for large HgTe nanocrystals~\cite{Apretna2021}.

For Drude absorption, analyzing the energy relaxation times requires a knowledge of the temperature dependence of the absorption coefficient, Eq.~\eqref{Is_Drude}, for the electron gas in quantizing magnetic fields. This remains the subject of an independent study.

Now, we discuss the data at intensities higher than  10000~W/cm$^2$. The inset in Fig.~\ref{Fig_Int_015}(a) shows that at the highest intensities, the sublinear dependence transits to a superlinear one. At first glance, this surprising effect can be attributed the generation of additional free carriers due to THz impact ionization of impurity states or transitions across the band gap~\cite{Ganichev2005}. A radiation induced increase in the free carrier density increases of the Drude-like absorption  and, consequently, the MPGE current magnitude. 
Light-induced  impact ionization, first observed in bulk InSb~\cite{Ganichev1984}, has recently been detected in HgTe based systems~\cite{Hubmann2019}. A key signature of this type of impact ionization, induced by the THz electric field, is that the angular radiation frequency $\omega=2\pi f$ is much higher than the reciprocal momentum relaxation time. Thus, impact ionization occurs solely due to collisions in the presence of a high-frequency electric field~\cite{Ganichev1984,Hubmann2019,Ganichev2005}.
The probability of impact ionization is proportional to the  exponential function, $\exp(-E_0^2/E^2)$, 
where $E$ is the radiation electric field amplitude and $E_0$ is a characteristic field parameter. Therefore,  the corresponding increase of the MPGE current varies with the radiation intensity as
\begin{equation}
	\label{impact}
J^{\rm MPGE}_{impact} = C \exp(-{I_0\over I}),
\end{equation}
where $C$ is the MPGE amplitude and ${I_0  =n_\omega\abs{E_0}^2/(2Z_0)}$. 
Using the sum of Eqs.~\eqref{015fit} and~\eqref{impact} with $C=1730$~$\mu$A and $I_0=38000$~W/cm$^2$, we successfully fit the entire data set across seven orders of magnitude in intensity, see the solid line in Fig.~\ref{Fig_Int_015}(a) and the inset. Notably,  the THz impact ionization process in  Cd$_{x}$Hg$_{1-x}$Te films has not been studied so far and is of independent interest. Investigating this phenomenon requires substantially higher radiation intensities and remains the subject of a separate study.

\subsection{Structures with noninverted band order ($x=0.22$)}

Now, we turn to the analysis of the data obtained for the structure with $x=0.22$, which is characterized by a  noninverted band order. We begin with the data obtained at 4.2~K and $B_i=0.29$~T. The inset in Fig.~\ref{Fig2_B_022}(a) shows that under these conditions the MPGE current arises from the superposition of two resonant absorption processes, the cyclotron resonance (CR) and ionization of impurities ($\cal{R}^{\rm imp}$), which were studied in Ref.~\cite{Moldavskaya2024}. Similar to sample~A, indirect Drude optical transitions provide a third contribution.  The CR MPGE is only possible in the CRA configuration, i.e. for $\alpha =0$, and manifests as a strong resonance peak at $B=B_i$ in Fig.~\ref{Fig2_B_022}(a). The MPGE due to Drude absorption and ionization of impurities can be excited in both CRA and CRP configurations. The Drude processes result in the smooth, nonmonotonic behavior of the MPGE, as shown by the black solid lines in Figs.~\ref{Fig2_B_022}(a) and~\ref{Fig2_B_022}(b). The MPGE caused by the impurity ionization is insensitive to the polarization and appears as a peak above the Drude background at $B_i=0.29$~T for the CRP configuration, see Fig.~\ref{Fig2_B_022}(b) where the CR vanishes, as well as a contribution to the resonance signal in the CRA configuration. Notably,  both resonances have their maxima at almost the same magnetic field, see the inset in Fig.~\ref{Fig2_B_022}(a).

\begin{table}[t]
	\centering
	\caption{Parameters used for fitting of the data for sample B ($x=0.22$) in Fig.~\ref{Fig2_Int_022} , based on 	Eqs.~\eqref{CRA015} and~\eqref{CRP015}.
		The fitting parameters are given for the CR, interband optical transitions  and Drude mechanisms of the MPGE.
	}
	\begin{tabularx}{0.5\textwidth}{XXXXXX}
		\toprule[0.05cm]\addlinespace[0.2cm]
		 & Transitions  & A$^i$ ($\mu$A) & $I_s^i$ (W/cm$^2$) \\
		\midrule[0.025cm]\addlinespace[0.1cm]
		CRA & CR           & 1443      & 0.0045               \\
		& Impurity     &           &                    \\
		& ionization   & 113      & 0.25                \\
		& Drude        & 9      & 62              \\	
		\midrule[0.025cm]\addlinespace[0.1cm]
		CRP & Impurity    &            &                    \\
		& ionization  & 41       & 0.25          		\\
		& Drude       & 6      & 90                \\
		\addlinespace[0.1cm]\bottomrule[0.05cm]
	\end{tabularx}
	\label{Tab2}
\end{table}

Following the theory developed above, see Eqs.~\eqref{K} and~\eqref{KDr}, and considering that the CR can only be excited in CRA configuration, the intensity dependence of the total MPGE current is given by
\begin{equation}
	\label{CRA015}
J^{\rm MPGE}_{\alpha=0} 
= {A^{\rm CR}I\over 1+ I/I_s^{\rm CR}} + {A^{\rm imp}_{\rm CRA}I\over 1+ I/I_{s,{\rm CRA}}^{\rm imp}} + {A^{\rm Dr}_{\rm CRA}I\over 1+ I/I_{s,{\rm CRA}}^{\rm Dr}},
\end{equation}
for CRA, and 
\begin{equation}
		\label{CRP015}
J^{\rm MPGE}_{\alpha=90^\circ} = {A^{\rm imp}_{\rm CRP}I\over 1+ I/I_{s,{\rm CRP}}^{\rm imp}} + {A^{\rm Dr}_{\rm CRP}I\over 1+ I/I_{s,{\rm CRP}}^{\rm Dr}}
\end{equation}
for CRP configurations. Here $A^{i}_j$ with $i={\rm CR, imp, Dr}$ or $j = {\rm CRA, CRP}$ are the slopes of the MPGE current intensity dependencies in the linear regime for the corresponding optical transitions, and $I_{s,j}^{i}$ are the saturation intensities.

The fits of the data using Eqs.~\eqref{CRA015} and~\eqref{CRP015} are shown in Fig.~\ref{Fig2_Int_022}. As in the previous section, we fit the data by considering the photocurrent behavior in different intensity ranges -- first at low intensities, 
then at the moderate intensities, 
and finally at high intensities, for the latter see the inset in Fig.~\ref{Fig2_Int_022}.
This procedure allows us to use only one fitting parameter for each fitting step. Figure~\ref{Fig2_Int_022} demonstrates that for radiation intensities $I < 3000$~W/cm$^2$ Eqs.~\eqref{CRA015} and~\eqref{CRP015} accurately describe all the data. The corresponding fitting curves for the individual mechanisms are shown by dotted and dash-dotted lines. The obtained fitting parameters are given in Table~\ref{Tab2}, revealing that $A^{\rm CR} \gg A^{\rm imp} \gg A^{\rm Dr}$, and $I_s^{\rm CR} \ll I_s^{\rm imp} \ll I_s^{\rm Dr}$. Similar to the results obtained for structures with inverted bands, the Drude mechanism of the MPGE exhibits the smallest amplitude and the largest saturation intensities. In contrast, the CR mechanism of the MPGE has the greatest magnitude and the lowest saturation intensity.

We estimate the relaxation time  $\tau^{\rm CR}$ from the intensity saturation data. According to Eq.~\eqref{IsCR_A}, for $\hbar\omega = 4.4$~meV, $m=0.008\:m_0$, $\tau_p=1.1$~ps, $n_\omega =3$ and
$I_s^{\rm CR,A} =0.45\times 10^{-2}$~W/cm$^2$, we obtain $\tau^{\rm CR} \approx 16$~ns. This value is of the same order of magnitude as the  energy relaxation times obtained for CR  in $n$-GaAs at liquid helium
temperature~\cite{Bluyssen1980,Weispfenning1985}.

An increase in temperature leads to the thermal ionization of impurity states, resulting in the MPGE current due to Drude absorption. This results in a nonresonant magnetic field dependence, as shown in Fig.~\ref{Fig_109K_022}. The difference between the curves for azimuth angles $\alpha =0$ and $\alpha =90^\circ$, observed at $B\gtrsim 0.7$~T, is caused by the polarization dependent mechanism of the MPGE, already addressed above~\cite{Foot_pol_dep}. An increase in temperature leads to a decrease in energy relaxation times due to  phonon scattering and a smaller deviation of electron temperature from the lattice temperature. Consequently, the saturation intensity $I_s^{\rm Dr}$ increases. Our data shows that at high $T$, the photocurrent is linearly dependent on $I$ over the studied intensity range, see Fig.~\ref{Fig_109K_022}(b).

The inset in Fig.~\ref{Fig2_Int_022} shows that for $I > 3000$~W/cm$^2$ a transition from sublinear to superlinear intensity dependence is observed. This is attributed to the \textit{light} impact ionization, as  discussed in the previous section for sample A.  Fitting the data using the sum of Eq.~\eqref{impact} and Eqs.~\eqref{CRA015},~\eqref{CRP015} provides a good description of the entire radiation intensity range, as shown by the solid curves in Fig.~\ref{Fig2_Int_022}. The fitting parameters for sample~B in  Eq.~\eqref{impact}  are $C=536$~$\mu$A and $I_0=6500$~W/cm$^2$.

\section{Summary}

Our experimental data and developed theory show that the MPGE current excited by intense terahertz radiation exhibits a complex nonlinear intensity dependence. The MPGE nonlinearity has been studied at liquid helium temperature and for magnetic field strengths where it arises from the superposition of resonant and nonresonant contributions. By analyzing the nonlinearities in both, the inverted and non-inverted band structures, we found that in both cases, they are caused by bleaching of radiation absorption. In the inverted band structures, observed and described the saturation of CR, resonant ionization of impurities, and nonresonant Drude-like absorption. In structures with a non-inverted band order, we studied MPGE arising from the resonant interband transitions between Landau levels and the Drude mechanism. The developed theory accurately describes the complex intensity dependence obtained by varying the irradiance over seven orders of magnitude. Our analysis reveals significant differences in the saturation intensities for the individual mechanisms, allowing us to fit the data over a wide intensity range using a single fitting parameter while keeping the other parameters fixed.  The obtained characteristics provide valuable insight for the development of CdHgTe devices in the THz spectral range.

\section{Acknowledgments}
\label{acknow} 
We thank I.~Yachniuk and S. N. Danilov for fruitful discussions. The financial support of the Deutsche Forschungsgemeinschaft (DFG, German Research Foundation) via Project-ID 521083032 (Ga501/19) is gratefully acknowledged.  
We also acknowledge the Russian Science Foundation via Project 24-62-00010.

\bibliography{all_lib1.bib}

\end{document}